\documentclass[a4]{article}
\usepackage[left=2cm,right=2cm,top=2cm,bottom=2cm,bindingoffset=0cm]{geometry}
\usepackage{epsfig}
\usepackage{epstopdf}
\usepackage{multirow}
\newcommand{\Ref}[1]{(\ref{#1})}
\newcommand{\beao}{\begin{eqnarray*}}
\newcommand{\eeao}{\end{eqnarray*}}
\newcommand{\be}{\begin{equation}}
\newcommand{\ee}{\end{equation}}
\newcommand{\bea}{\begin{eqnarray}}
\newcommand{\eea}{\end{eqnarray}}
\newcommand{\beq}{\begin{eqnarray}}
\newcommand{\eeq}{\end{eqnarray}}
\newcommand{\nn}{\nonumber}

\begin{document}

\title{On plasmon contribution to the hot $A_0$ condensate}
\author{O. A. Borisenko 
\thanks{e-mail: Oleg@BITP.Kiev.ua}\\
{\small N.N. Bogolubov Institute for Theoretical Physics, 252143 Kiev, Ukraine}\\
\\
 V. V. Skalozub \thanks{e-mail:Skalozubv@daad-alumni.de}\\
{\small Oles Honchar Dnipro National University, 49010 Dnipro, Ukraine}}

\date{}
\maketitle.


%

%
\begin{abstract}
In $SU(2)$ gluodynamics, the Debye gluon contribution $W_D(A_0)$
to the effective action of the temporal gauge field component,
$A_0 = const$, at high temperature is calculated in the background
$R^{ext}_\xi $ gauge.  It is shown that at $A_0 \not = 0$ the
standard definition $k_0 = 0,~ |\vec{k}| \to 0$  corresponds to
long distance correlations for the  longitudinal in internal space
gluons. The transversal gluons become screened by the $A_0$
background field. Therefore  they give zero contributions and have to be
excluded from the correlation corrections. The  total effective action
accounting for the one-loop, two-loop and   correct  $W_D(A_0)$
satisfies Nielsen's identity that proves  gauge invariance of the
$A_0$ condensation phenomenon.
\end{abstract}

\section{Introduction}
Investigation of QCD high temperature phase - quark-gluon plasma
(QGP) - is a paramount problem nowadays. The order parameter of the
phase transition here is Polyakov's loop. In the imaginary
time formalism it is the  integral of the gluon field  component $A_0$
along an imaginary time direction contour. This integral
observable is not a solution to gluon field equations. Therefore instead a
related parameter, so-called $A_0$ condensate, $A_0$ = const,  is
also discussed. It is a constant part of the temporal gauge field
component.
In perturbation field theory, the $A_0 \not = 0$ has been determined in loop
expansion of an effective potential in two loop order
\cite{Anishetty}-\cite{Enqvist}. Different aspects of the condensation
are discussed in the literature. Numerous references can be found,
in particular, in review paper \cite{Borisenko95}. This classical field would be very essential for
phenomenology. It is relatively simple to take it into account in actual  calculations. It looks as 
an imaginary chemical potential in finite temperature field theory. 
Influence of
$A_0$ on various processes has also been discussed (see references
in recent papers \cite{Bordag2019}, \cite{Skalozub2019}).

In \cite{Belyaev91} - \cite{Sawayanagi92} the gauge fixing
independence  (and hence a gauge invariance) of $A_0 \not = 0$
has been called in question. In particular, it was stated that the
contributions of the plasmon diagrams to the effective action
$W_D(A_0)$, which describe the long-range correlation
corrections to the one-loop effective potential $W^{(1)}(A_0)$,
cancel the part of the two-loop one, $W^{(2)}(A_0)$,  and the
$A_0 $ = 0 must be detected. However in contrast, in  \cite{Skalozub94}, \cite{Skalozub94b},
\cite{Skalozub94a} it has been shown that for the sum of the
one-loop plus two-loop effective action  the result $A_0 \not =
0$ followed. Just for this case Nieslen's identity holds. This fact, in
accordance with a general theory (see, for example
\cite{Borisenko95}),  means that the $A_0$ condensation  is 
gauge invariant phenomenon which is realized at two-loop level. The
contradiction of these conclusions is obvious. 

In the present paper we investigate the role of
plasmon diagrams  in more details. To realize that, using the
SU(2) gluodynamics as an example, we calculate the plasmon
contribution in a general relativistic $R^{ext}_\xi $ gauge. We
show that at  $A_0 \not = 0$ the screening at low momenta of
the transversal  color field modes  takes place. So that, they
do not give contributions to the effective potential of
correlation corrections. This is in contrast to the calculation
procedures applied in the Feynman gauge in \cite{Belyaev91}, \cite{Sawayanagi92}.
Hence   the cause for the discrepancy of the  results 
\cite{Belyaev91}, \cite{Sawayanagi92} and   \cite{Borisenko95}
becomes  clear. We derive the  correct expression for the
plasmon contributions $W^{(3)}_D (A_0)$ (see \Ref{WD3}) which is gauge fixing
independent.  Nielsen's identity holds for the total effective
action  $W^{(tot)}(A_0) = W^{(1)}(A_0)$ +$W^{(3)}_D (A_0)$ +
$W^{(2)}(A_0)$. That proves the gauge invariance of the $A_0 $
condensate.

  In next section, we present a  general theory of  
 investigations   and some previous results  necessary for what follows.
 In section 3, we carry out actual calculations.    The
discussion of the results obtained  and  the difference between the cases  of the $SU(2)$
and $S(3)$ gluodynamics are given in the last section.

\section{Consideration at two-loop order}

Let us consider   $SU(2)$ gluodynamics in the Euclidean space
time embeded  in the background field $\bar{A}^a_\mu =  A_0
\delta_{\mu 0 } \delta^{a 3} =const $ described by the
Lagrangian
\be \label{L0} L = \frac{1}{4} (G^a_{\mu \nu})^2 + \frac{1}{2 \xi}  [(\bar{D}_\mu A_\mu)^a]^2 - \tilde{C} \bar{D}_\mu D_\mu C. \ee
The gauge field potential $A_\mu^a =   Q_\mu^a +  \bar{A}_\mu^a$ is
decomposed  in quantum and classical parts. The covariant
derivative in Eq. \Ref{L0} is $(\bar{D}_\mu A_\mu)^{ab} =
\partial_\mu \delta^{ab}
 -  g \epsilon^{abc}  \bar{A}_\mu^c,~ G_{\mu\nu}^a =  \bar{D}_\mu Q_\nu^a - \bar{D}_\nu Q_\mu^a - Q_\mu^b Q_\nu^c$, g is a coupling constant, internal index $a = 1, 2, 3.$  The Lagrangian of ghost fields $\bar{C}, C$ is determined by the backgraund covariant derivative $\bar{D}_\mu (\bar{A}) $ and the total one $D_\mu (\bar{A} + Q. $ As in  \cite{Enqvist}, \cite{Skalozub94} we introduce the "charged basis" of fields:
\bea \label{chbasis}A^0_\mu &=& A^3_\mu, ~~A^{\pm}_\mu = \frac{1}{\sqrt{2}} ( A^1_\mu \pm i A^2_\mu ), \nn \\
           C^0 &=& C^3, ~~C^{\pm} = \frac{1}{\sqrt{2}} ( C^1_ \pm i C^2 ) .\eea
In this basis a scalar product is $x^a y^a = x^+ y^- + x^- y^+
+ x^0 y^0$,  and the structure constants are: $\epsilon^{abc} =
1 $ for a = "+", b = "-", c = "0". Feynman rules are the usual
ones for the theory at finite temperature with modification: in
the background field $\bar{A}^a_\mu $ a sum over frequencies $
\frac{1}{\beta} \Sigma (k_0 = 2 \pi /\beta)$ should be replaced
by $ \frac{1}{\beta} \Sigma (k_0 = 2 \pi /\beta \pm g
\bar{A}^a_0)$ in all loops of the fields $Q^{\pm}_\mu,
C^{\pm}$. This frequency shift must be done not only in
propagators but also three particle vertex \cite{Belyaev91}.
The effective action $W(A_0)$ is given as a functional integral
over fields with a compact imaginary time direction $0 \le  t
\le 1/T = \beta$:
\be \label{W} \exp[- W(\bar{A}_0) V T] = N \int DQ DC D\tilde{C} \exp \bigl [ - \int\limits^\beta_0  d \tau \int d^3 x (L - Q J) \bigr ], \ee
where N is T-independent normalization factor, V is a space
volume, J is an external source.
The effective action up to two-loop order reads:
\bea \label{efact} W(x) &=& W^{(1)}(x) + W^{(2)}(x) , \nn \\
 \beta^4  W^{(1)}(x) &=&  \frac{2}{3} \pi^2 [ B_4 (0) + 2 B_4 (x/2)]; \nn \\
 \beta^4  W^{(2)}(x) &= & \frac{1}{2} g^2 [ B_2^2(x/2) + 2 B_2 (x/2) B_2 (0) ] \nn \\
&+& \frac{2}{3} g^2 (1 - \xi ) B_3 (x/2) B_1 (x/2 ),\eea
where we introduce the dimensionless variable $x = \frac{g A_0}{\pi T}$ and
\bea \label{Bernoulli} B_1 (x) &=& x - \frac{1}{2} \epsilon (x), \nn \\
 B_2 (x) &=& x^2 - |x| + \frac{1}{6}, \nn \\
 B_3 (x) &=& x^3  -  \frac{3}{2} \epsilon (x) x^2   + \frac{1}{2} x, \nn \\
 B_4 (x) &=& x^4  - 2 |x|^3 + x^2 - \frac{1}{30}, \eea
are the Bernoulli polynomials, $\epsilon (x) = x/|x|$.  For $\xi = 1$ it has been calculated in \cite{Belyaev91}  (for SU(3) theory see \cite{Enqvist}, \cite{Skalozub94}).

 As we see, W(x) is $\xi$-dependent. This point served as an origin
for doubts in the gauge invariance of the gluon field
condensation phenomenon. As we mentioned in Introduction,
this problem has been solved within the Nielsen identity method in
\cite{Skalozub94} - \cite{Skalozub94a}. So here we restrict
ourselves to considering the plasmon contribution for $\xi$ to
be an arbitrary number.
\section{Plasmon contribution}
To be consistent, first we calculate the plasmon contribution in a way developed in  \cite{Belyaev91}, \cite{Sawayanagi92} for the value of $\xi = 1$. As it is well known \cite{Kalashnikov84}, \cite{Bellac96}, the plasmon contribution to the effective action $W_D$ is to be properly accounted for by suming the ring diagrams with leading infrared singularities, which present in  propagator $\sim 1/k^2$. To do that the difference between the infrared limit of the one-loop polarization tensors at finite temperature and the  zero temperature ones  for all the gluon fields,  $\Delta \pi$, should be computed.

Now,  let us calculate $W_D (A_0)$. By using the Feynman rules
described above and taking into account the explicit form of
the gluon propagator in the basis \Ref{chbasis},
\bea \label{propagator} D^{a b}_{\mu\nu} &=& \frac{\delta_{\mu\nu} \delta^{a b}}{(k^a)^2} - (1 - \xi)  \frac{k^a_\mu k^b_\nu }{(k^a k^a)^2}, \nn\\
\frac{\delta^{a b}}{(k^a)^2} &=& \Bigl[ \frac{\delta^{+ -}}{(k^+)^2},~  \frac{\delta^{- +}}{(k^-)^2}, ~  \frac{\delta^{0 0}}{(k^0)^2} \Bigr], \eea
where $(k^{\pm})^2 = (k_0 \pm g \bar{A}_0)^2 + \vec{k}^2$, we obtain:
\be \label{WD} W_D (\bar{A}_0, \xi ) = -\frac{1}{2} \sum\limits_{k_0 = 2 \pi n T} \int \frac{d^3 k}{(2 \pi)^3}  \sum \limits_{a,b = (+,-, 0)} \frac{(\Pi^{a b}(0))^2 (D^{ b a}(k))^2}{1 - (\Pi^{a b}(0)) (D^{ b a}(k))},\ee
where $\Pi^{a b}(0)) = \Delta \pi^{a b}_{0
0}(\vec{k})|_{\vec{k} \to 0}$ is the asymptotic form of the
one-loop polarization tensor and limit $|\vec{k}| \to 0$ is to
be calculated in a way depending on the definition of "infrared
mass shell" at $T \not = 0, A_0 \not = 0$.

Now, we are going to calculate Eq.\Ref{WD} in three ways. First
was proposed in  \cite{Belyaev91},\cite{Sawayanagi92}.  Let us
consider the standard definition of $\Pi^{a b}$,
\be \label{Piab} \Pi^{a b} = \Delta \pi^{a b}_{0 0}(k) |_{k_0 = 0, |(\vec{k})|\to 0} = - m^2_D, \ee
where $m^2_D = \frac{2}{3} g^2 T^2$ is the Debye mass squared
\cite{Belyaev91} and all x-dependent  terms were omitted as in
\cite{Belyaev91} - \cite{Sawayanagi92}. Substituting expressions
 \Ref{propagator} and performing integration we get
\bea \label{WD1} W_D^{(1)}&=&\frac{g^2}{6 \beta^4} \Bigl[ -\frac{g}{6 \sqrt{\pi} \sigma} + \frac{1}{4 \sigma} \lambda^- (1 - \sigma)^2 \nn \\
&-& \frac{1}{4 \sigma} \lambda^+ (1 + \sigma)^2 + \frac{x}{2} (3 - \xi) \Bigr] + f(n \not = 0), \eea
where
\be \label{sigmalambda} \sigma = \Bigl( 1 + \frac{6 (1 - \xi) x^2 \pi^2}{g^2} \Bigr)^{1/2} ; ~ \lambda^{\pm} = \Bigl ( x^2 + \frac{g^2}{3 \pi^2} (1 \pm \sigma) \Bigr)^{1/2} \ee
and the explicit expression for $\Pi_{0 0}$ was used. As usually (see \cite{Belyaev91},\cite{Sawayanagi92}) only the zero mode, n = 0, is picked out and the nonstatic mode contributions are denoted as $ f(n \not = 0)$. As we see, $ W_D^{(1)}$ is $\xi$-dependent and for $\xi = 1$ it coincides with the results of \cite{Belyaev91},\cite{Sawayanagi92}. In contrast to statemnts of these papers, one can conclude that the applied procedure results in the gauge variant expression. It is easy to verify using Eqs.\Ref{efact},\Ref{WD1} that in the sum $W^{(2)} + W_D^{(1)} = W^{(1)}_{tot}$ the linear terms are cancelled, as in the Feynman gauge. So, folowing the idea of  \cite{Belyaev91},\cite{Sawayanagi92}, we would conclude that there is no $A_0$ condensation (in order $\sim g^2$).

However, the $\xi$-dependence of $ W_D^{(1)}$ my call a doubt in reliability of the latter conclusion. From theoretical grounds, it is well known that the plasmon contrbutions are gauge invariant either in QED or QCD \cite{Kalashnikov84}, \cite{Bellac96}. Next, $\xi$ dependent terms in  $ W_D^{(1)}$  have the order  $\sim g^2$. So, this functional must be taken into account together with \Ref{efact}. But the Nielsen identity holds for the fuctional \Ref{efact} alone \cite{Skalozub94}-\cite{Skalozub94a}. Hence it immediately follows, the properties described by the sum $ W^{(1)}_{tot}$ occur to be gauge dependent. In particular, this concerns the result $A_0 = 0$ of  \cite{Belyaev91},\cite{Sawayanagi92}.

It is not difficult to find the origin of the gauge dependence of $ W_D^{(1)}$. It comes out from the definition of the infrared mass shell in the case of  $A_0  \not =  0$ used in these papers. Really, as it is well known, the plasmon corrections sum up the singular infrared contributions of the tree-level propagators and have the order $g^3$  in coupling  constant. However, from Eq.\Ref{propagator} it follows that only for the longitudinal modes in the internal space (a = b = 0) the standard definition: $k_0 = 0, |\vec{k}| \to 0$ reproduces the infrared divergency of $D^{00}(k)$. At arbitrary $A_0  \not = 0$  for transversal modes ($a, b = \pm)$ this limit is a regular one.

To have a singular infrared contribution a slightly modified definition of infrared mass shell for transversal modes should be introduced: $(k_0  \pm g A_0) = 0, |\vec{k}| \to 0$.      With this definition used the infrared singularity of  the $D^{+ -}_{00} \sim  1/|\vec{k}|^2 $ is reproduced and its gauge invaiance is obvious. To incorporate this definition in Eq.\Ref{WD} it is necessary to calculate the components $\Pi^{+-}_{00}(A_0 \not = 0, k)$ of the one-loop polarization tensor.

Omitting standard one-loop calculations let us write down the final result,
\bea \label{pia0} &-& \Pi^{+-}_{00}|_{(k_0  \pm g A_0) = 0, |\vec{k}| \to 0} = 2 g^2 T^2 \cdot (B_2(0) + B_2(x/2)) \nn \\
&-& \Pi^{00}_{00}|_{ k_0  = 0, |\vec{k}| \to 0}= 4 g^2 T^2  B_2(x/2)). \eea
Substituting expressions \Ref{pia0} in Eq.\Ref{WD} and integrating over momentum space, we obtain
\be \label{WD2} W_D^{(2)}(A_0) = - \frac{2 g^3 T^4}{3 \pi} \Bigl [ B^{3/2}_2 (x/2) + 2 [\frac{1}{2}(B_2 (x/2) + B_2 (0) )]^{3/2} \Bigr]. \ee
This nonanalytic term has the order $\sim g^3$ and is gauge
invariant,  as it should be for plasmon diagram corrections
\cite{Kalashnikov84}.

Now, another question  arises: How the condition $(k_0  \pm g
A_0) = 0$ could be implimented in the  imaginary time formalism?
In fact, it is not possible at arbitrary values of $A_0$
because of the $ k_0 = 2 \pi T n, n = 0, \pm 1, \pm 2,...$. The
only way is $g A_0 = 0$ for n = 0, $g A_0 = 2 \pi T$ for n = -
1,  $g A_0 = - 2 \pi T$ for n = + 1. These are the values
corresponding to the $Z(2)$ symmetry which takes place for
$SU(2) $ gauge fields for unbroken   symmetry of the Lagrangian
\Ref{L0}  at finite temperature.  For these values of $A_0$ the
values of $B_2(x/2) = 1/6$ and the squared bracket in Eq.
\Ref{WD2} is the fixed number. This case corresponds to the
effective action including the one-loop plus $W_D^{(2)}(A_0)$
which describes the Z(2) phases without symmetry breaking. This
is the exact meaning of the second considered definition of the
infrared mass shell.

 However, to investigate the spontaneous symmetry breaking we have to
  put $A_0$ value to be arbitrary and determine it from the minimum
   condition of the effective action. Since for this case the symmetry is broken,
   no infrared singularities present for the transversal propagators in Eq.\Ref{propagator}.
    So that we have to drop the contribution of these modes in Eq.\Ref{WD2}, which comes
    from polarization tensor \Ref{pia0}. Thus, the correct  expression for the plasmon contribution is
\be \label{WD3} W_D^{(3)}(A_0) = - \frac{2 g^3 T^4}{3 \pi}  B^{3/2}_2 (x/2)  . \ee
It comes from neutral gluon components of the internal space
and obviously is gauge invariant. For this case the Nielsen
identity  holds and the effective action
$W_{tot} = W^{(1)}(A_0) + W^{(2)}(A_0) + W_D^{(3)}(A_0)$ has a
non-trivial minimum $A_0  \not =  0$. Thus, in accordance with
general principles of this approach (see \cite{Skalozub94},
\cite{Borisenko95}),  one has to conclude that the  $A_0$
condensation takes place at two-loop level and is a gauge
invariant phenomenon.
\section{Discussion}
In the present paper we analyzed the role of the plasmon
contributions to the  effective action of the gluon condensate $A_0 = const$
at high temperature. Our main result is two-fold. First we have
shown explicitly that the  conclusion of \cite{Belyaev91},
\cite{ Sawayanagi92} derived from the effective action
$W^{(1)}_{tot} (A_0, \xi = 1)$ (Eqs.\Ref{efact}, \Ref{WD1}) is
inconsistent with the Nielsen identity and so occurs to be
gauge non-invariant. We have calculated the gauge independent
plasmon contribution $W^{(2)}_D$ \Ref{WD2} which takes into consideration
the special definition of the infrared mass shell $k_0 \pm g
A_0 = 0, |\vec{k}| \to 0$ for the transversal internal space
gluons, which corresponds to the Z(2) unbroken symmetry. This
possibility has also been mentioned in \cite{Enqvist}.  But
for broken symmetry at high temperature these modes become
massive and should be excluded from the long range corrections.
As a result, the only longitudinal in color space gluon modes
occur to be long range and contribute to the plasmon effective action.
It is gauge fixing independent and satisfies  the Nielsen
identity. This effective action has a nontrivial minimum that
means the gauge invariance of the gluon condensation phenomenon
as a whole. In this approach the $\xi$-dependence of the
minimum position simply means that there is a set of special
unknown this moment diagrams which contribution  cancels the
non-invariant terms. The cancelation does not change the
minimum value of the effective action as well as other
characteristics of particles.

One of applications of the results obtained is the early Universe before the electroweak 
 phase transition. At high temperatures, the $SU(2)_{EW} \times U(1)_Y$ symmetry is 
 restored and the W and Z bosons as well as photons convert into charged and neutral
non-Abelian and Abelian gauge fields belonging to the  initial gauge groups. For the former fields, all
 the results obtained are relevant. That means the presence of the $A_0^{w.} \not = 0$  condensate generated
 in the weak sector of the Standard Model. The details   on this phenomenon will
 be present in other publication.
       The  description of the plasmon  contributions was not given in the review \cite{Borisenko95}
 or elsewhere else.  So, the present paper removes this shortcoming.

To complete we would like to note that considered  SU(2) gluodynamics
differs a little from the SU(3) case. In the latter one, two
background fields $A_0^3$ and $A_0^8$ corresponding to the commuting generators
$\frac{\lambda^3}{2}$ and $\frac{\lambda^8}{2}$
 are expected to be
generated.  In principle, some  combinations of these fields could
become massless. It may happen after the diagonalization of the
non-diagonal matrix of charged gluon fields appearing from
the one-loop polarization tensors entering Eq.\Ref{WD}. This possibility is
accepted in \cite{Enqvist}. It is realized in case when
both of condensed fields are nonzero. However, as it is shown
in \cite{Skalozub94}, \cite{Skalozub94b}, \cite{Skalozub94a}, \cite{Borisenko95},
at two-loop level only the condensate $A_0^3 \not = 0 $ is
generated. Again, the one- plus two- loop effective action satisfies the Nielsen identity.   There are no massless (or unstable)
charged modes at high temperature and the situation in SU(3)
gluodynamics  is similar to the investigated case.

\end{document}